\documentclass[12pt]{iopart}

\newif\ifpdf
\ifx\pdfoutput\undefined
\pdffalse 
\else
\pdfoutput=1 
\pdftrue
\fi
\ifpdf
\usepackage[pdftex]{graphicx}
\usepackage[pdftex]{color}
\usepackage{thumbpdf}
\usepackage{hyperref}
\hypersetup{
colorlinks=true,
linkcolor=blue,
pagecolor=white,
pdftitle={
Dynamical density functional theory for interacting Brownian particles:
stochastic or deterministic?
},
pdfauthor={Andrew J. Archer and Markus Rauscher},
pdfsubject={Paper to be submitted to J. Phys. A: Math. Gen.},
pdfkeywords={DDFT, Brownian}
}
\else
\usepackage[dvips]{graphicx}
\usepackage[dvips]{color}
\newcommand{\href}[2]{#2}
\fi

\usepackage{iopams}
\usepackage{bm}
\usepackage{times}

\newcommand{\vct}[1]{\bi{#1}}

\newcommand{\grad}{\bm{\nabla}}

\newcommand{\cdensity}{\bar{\rho}}
\newcommand{\adensity}{\rho}
\newcommand{\corrfunc}{\rho^{(2)}}
\newcommand{\eqcorrfunc}{\rho^{(2)}_{eq}}
\newcommand{\dcorr}{c^{(1)}}
\newcommand{\mdensity}{\hat{\rho}}
\newcommand{\pdensity}{w}
\newcommand{\Pdensity}{W}
\newcommand{\Cdensity}{\bar{W}}
\newcommand{\intpot}{V}
\newcommand{\intforce}{\vct{F}}
\newcommand{\expot}{\Phi}
\newcommand{\exforce}{\vct{G}}
\newcommand{\Hamil}{\mathcal{H}}
\newcommand{\free}{\mathcal{F}}
\newcommand{\freeex}{\free_{\!\!{ex}}}
\newcommand{\kernel}{K}

\begin{document}

\title[Time dependent density functional theory]
{Dynamical density functional theory for interacting Brownian particles:
stochastic or deterministic?}
\date{\today}

\author{Andrew J. Archer\dag\ and Markus Rauscher\ddag\P}
\address{\dag\ H.~H.~Wills Physics Laboratory, University of Bristol,
Bristol BS8 1TL, UK}
\address{\ddag\ Max-Planck-Institut f\"{u}r
Metallforschung, Heisenbergstr.\ 3, 70569 Stuttgart, Germany}
\address{\P\ ITAP,
Universit\"{a}t Stuttgart,  Pfaffenwaldring 57, 70569 Stuttgart, Germany}
\ead{\href{mailto:andrew.archer@bristol.ac.uk}{andrew.archer@bristol.ac.uk}}
\ead{\href{mailto:rauscher@mf.mpg.de}{rauscher@mf.mpg.de}}
\begin{abstract}
We aim to clarify confusions in the literature as to whether or not 
dynamical density functional theories for the one-body density of a
classical Brownian fluid should contain a stochastic noise term. We point out
that a stochastic as well as a
deterministic equation of motion for the density distribution
can be justified, depending on how the fluid one-body
density is defined -- i.e.\ whether it is an ensemble averaged
density distribution
or a spatially and/or temporally coarse grained density distribution.
\end{abstract}

\submitto{\JPA}


\section{Introduction}
\label{sec:intro}

In equilibrium statistical mechanics, one can
prove that the grand canonical free energy of the system can be written as a
functional of the one-body
density only \cite{evans79}\/. The density distribution which minimizes the
grand potential functional is the equilibrium density distribution. This
statement is the basis of the equilibrium
density functional theory for classical
fluids which has been used with great success to describe a variety of
inhomogeneous fluid phenomena \cite{evans92}\/. Out of equilibrium
there is no such rigorous principle. However, macroscopically one
can find phenomenological equations for the time evolution 
which are based on macroscopic quantities only, e.g.\ the diffusion
equation, the heat transport equation and the Navier-Stokes equations for
hydrodynamics. 

In perhaps one of the most simple microscopic cases, a system of interacting
Brownian particles, a number of equations for the time evolution of the particle
density have been proposed which aim
to incorporate results from equilibrium density functional theory.
These equations are usually referred to as dynamical density
functional theory (DDFT) or time dependent density functional theory. 
Most of the equations take a rather similar form but there has been some
controversy as to whether the time evolution ought to be stochastic
\cite{munakata89,kirkpatrick89,dean96,kawasaki94,frusawa00} 
or deterministic \cite{marconi99,marconi00,archer04} in
nature. In reference \cite{frusawa00} the authors attempt to address these issues.
However, we believe further clarifications are necessary.
The reason for the controversy is that one has to
state precisely what type of density distribution
one is talking about, namely whether
one speaks of the instantaneous density $\mdensity(\vct{r},t)$ at time $t$, an
ensemble averaged density $\adensity(\vct{r},t)$, or a spatially and/or
temporally coarse grained density $\cdensity(\vct{r},t)$\/. We will give
proper definitions for these densities later.
The time evolution equation for $\mdensity(\vct{r},t)$ is discussed in
references
\cite{dean96,kawasaki94,frusawa00,marconi99} and for Brownian particles
this is, of
course, a stochastic equation. $\adensity(\vct{r},t)$ results from taking an
ensemble average over the stochastic noise and is therefore uniquely valued at
time $t$; the equation governing its dynamics must therefore
be deterministic -- this is the approach
in references \cite{marconi99,marconi00,archer04}. The equation governing the
dynamics of $\cdensity(\vct{r},t)$, the coarse grained density, will, of
course, still contain a noise term.
The main objective of this paper is to clarify the relationships between
these three approaches and to discuss the similarities
and differences. In order to visualize better the connections between the DDFTs
for $\mdensity(\vct{r},t)$, $\adensity(\vct{r},t)$ and $\cdensity(\vct{r},t)$,
we employ the flow chart in \fref{figure}\/.

This paper proceeds as follows:
In \sref{sec:brown} we formulate the dynamics of a system of interacting
Brownian particles. We derive a deterministic 
DDFT for the ensemble (or noise) averaged density in \sref{sec:ensemble}\/.
In \sref{sec:coarse} we formulate the stochastic DDFT for the coarse grained
density and finally, in \sref{sec:conclusion}, we discuss the results.

\section{Brownian dynamics of interacting particles}
\label{sec:brown}

\begin{figure}
\begin{center}
\includegraphics{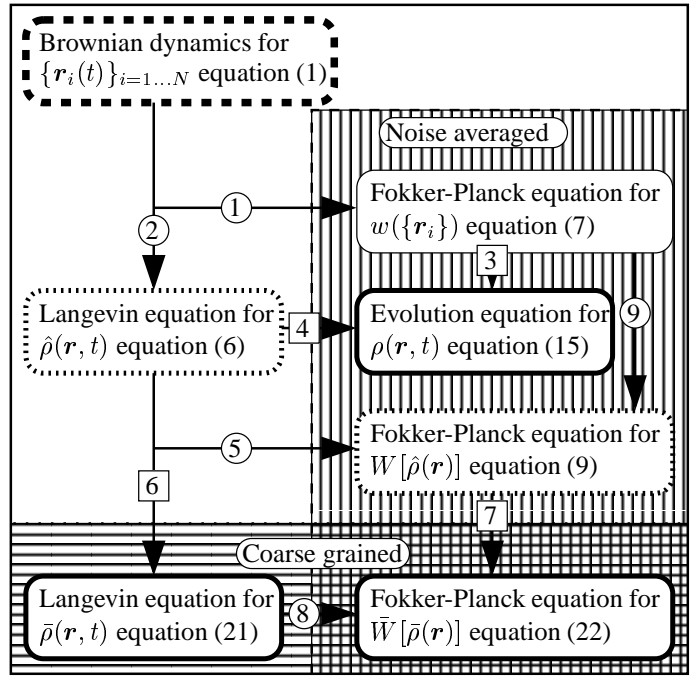}
\end{center}
\caption{\label{figure} Flow chart outlining the various possible routes for
deriving a DDFT for
pairwise interacting Brownian particles. The starting point for all attempts to
derive a DDFT is equation \eref{eq:brownian} (thick dashed line
box at the top)\/. The
end point is either a theory for the coarse grained density
$\cdensity(\vct{r},t)$ (Langevin or Fokker-Planck equation, thick solid
outlined boxes at the
bottom) or, for the ensemble averaged density $\adensity(\vct{r},t)$,
a deterministic equation, (thick solid
outlined box on right)\/. The Langevin and
Fokker-Planck equation for the microscopic density $\mdensity(\vct{r},t)$
are only intermediate results (dotted boxes)\/. Steps 1, 2, 5, 8, and 9 are
exact (in circles) whereas steps 3, 4, 6, and 7 involve
approximations (in square boxes). Ensemble averaged theories are located in
the vertically striped area while coarse grained theories are located
in the horizontally striped area.
}
\end{figure}

The starting point for our considerations is the Langevin equation
for a system of $N$ identical Brownian particles with
positions $\{\vct{r}_i(t)\}_{i=1\dots N}$
(see the dashed box in \fref{figure}).
The particles interact pairwise via the (pair) potential $\intpot(r)$. In
addition, there is a one-body external potential $\expot(\vct{r}_i,t)$\/.
This equation reads: 
\begin{equation}
\label{eq:brownian}
\frac{\rmd \vct{r}_i}{\rmd t} = 
\sum\limits_{j=1\dots N} \intforce(\vct{r}_i - \vct{r}_j) 
+ \exforce(\vct{r}_i,t)
+ \bfeta_i(t),
\end{equation}
where $\intforce(\vct{r}_i - \vct{r}_j) = -\grad_{\! i}
\intpot(|\vct{r}_i - \vct{r}_j|)$ is the interaction force between a pair of
particles and $\exforce(\vct{r}_i,t)= -\grad_{\! i}
\expot(\vct{r}_i,t)$ is the external force on particle $i$\/.
The uncorrelated Gaussian random force on each particle
$\{\bfeta_i(t)\}_{i=1\dots N}$  is fully characterized by
\begin{equation}
\label{eq:bnoise}
\langle \bfeta_i(t)\rangle =0 \quad \mbox{and} \quad 
\langle \eta_i^\ell (t)\,\eta_j^m(t') \rangle =
2\,T\,\delta_{ij}\,\delta^{\ell m} \,\delta(t-t').
\end{equation}
The particles are labelled with lower roman indices and the spatial
directions (i.e., $x$, $y$, and $z$ in three dimensions) are labelled with
upper indices. A possible prefactor (mobility constant)
in front of the deterministic part of
\eref{eq:brownian} can be eliminated via the Einstein relation and a
corresponding rescaling of time. $T$ is
the temperature.
Note that the noise term in \eref{eq:brownian} is additive, i.e.\ it does not
have a prefactor which depends on the particle position. In this case, the
Ito and the Stratonovich stochastic calculus are equivalent \cite{risken84}\/. 

In the next step, we define the instantaneous microscopic density
$\mdensity(\vct{r},t;\{\vct{r}_i(t)\}
) = \sum_{i=1}^N \delta(\vct{r}_i(t)-\vct{r})$\/.
This density depends implicitly on the actual positions of all particles
$\{\vct{r}_i(t)\}$ and it should  be considered a
density `operator' rather than a measurable particle density.
Using Ito stochastic calculus, one can show that
the time evolution of this sum of delta functions is given by
(see references \cite[equation~(8)]{marconi99} or \cite[equation~(17)]{dean96}):
\begin{eqnarray}
\frac{\partial \mdensity(\vct{r},t)}{\partial t} = 
\grad \cdot \Bigg[ 
-\mdensity(\vct{r},t)\,\exforce(\vct{r},t) -
\mdensity(\vct{r},t) \, \int \mdensity
(\vct{r}',t)\,\intforce(\vct{r}-\vct{r}')\,\rmd^3 r' \nonumber\\+
T\,\grad \mdensity(\vct{r},t) +
\sqrt{\mdensity(\vct{r},t)}\,\bm{\xi}(\vct{r},t)\Bigg],
\label{eq:microevolution}
\end{eqnarray}
i.e., an equation which, at a first glance, does not depend upon the individual
particle positions. This step corresponds to path~2,
which leads to the left dotted box in \fref{figure}\/. 
The random forces on the particles  $\bfeta_i(t)$  
have been replaced by a Gaussian random field $\bm{\xi}(\vct{r},t)$ with
\begin{equation}
\langle \bm{\xi}(\vct{r},t) \rangle = 0 \quad \mbox{and}\quad 
\langle \xi^\ell(\vct{r},t)\,\xi^m(\vct{r}',t') \rangle = 
2\,T\,\delta^{\ell m} \,\delta(\vct{r}-\vct{r}')\,\delta(t-t').
\end{equation}
Introducing the `energy' functional $\Hamil[\mdensity]$ \cite{frusawa00}
\begin{eqnarray}
\label{eq:hamil}
\Hamil[\mdensity] = \int \Bigg\{
T\,\mdensity(\vct{r},t)\,\left[\ln \Lambda^3\mdensity(\vct{r},t)-1
\right]
+\mdensity(\vct{r},t)\,\expot(\vct{r},t)\nonumber\\
+ \frac{1}{2}\,
\int \mdensity(\vct{r},t)\,\intpot(|\vct{r}-\vct{r}'|)\,\mdensity(\vct{r}',t)\,
\rmd^3 r' 
\Bigg\}\,\rmd^3 r,
\end{eqnarray}
where $\Lambda$ is the de Broglie wavelength,
\eref{eq:microevolution} can be written as 
\begin{equation}
\frac{\partial \mdensity(\vct{r},t)}{\partial t}  = 
\grad\cdot\left[ \mdensity(\vct{r},t) \,\grad \frac{\delta
\Hamil[\mdensity]}{\delta \mdensity(\vct{r},t)} +
\sqrt{\mdensity(\vct{r},t)}\,\bm{\xi}(\vct{r},t)\right].
\label{eq:microevolution2}
\end{equation}
In contrast to the noise in equation \eref{eq:brownian}, the
noise in \eref{eq:microevolution2} is multiplicative. However, as shown in
\cite{rauscher04a}, for a conserved multiplicative noise, Ito and
Stratonovich calculus are equivalent and therefore
\eref{eq:microevolution2}, the Langevin equation for $\mdensity(\vct{r},t)$,
should have the same form when obtained by
Stratonovich calculus.

We denote the probability density for finding the particles in the system 
at time $t$ with the positions
$\{\vct{r}_{\ell}\}_{\ell=1\dots N}$
by $\pdensity (\vct{r}_1,\dots,\vct{r}_N,t)$\/. The time evolution of
this probability is given by the Fokker-Planck equation for
\eref{eq:brownian}, namely
\begin{equation}
\fl
\frac{\partial \pdensity(\{\vct{r}_{\ell}\},t)}{\partial t} = 
-\sum\limits_{i=1\dots N} \!\!\grad_{\! i}\,\cdot  
\left[ \sum\limits_{j=1\dots N}\!\!
\intforce(\vct{r}_i-\vct{r}_j) +\exforce(\vct{r}_i,t)
-T \,\grad_{\! i}\right] 
\pdensity(\{\vct{r}_{\ell}\},t).
\label{eq:positionfp}
\end{equation}
In \fref{figure} the derivation of \eref{eq:positionfp} is denoted by 
path~1\/. \Eref{eq:positionfp} is also referred to as the Smoluchowski
equation, namely the drift-diffusion equation
for a many particle system \cite{risken84,archer04}.

However, we are more interested in the particle density distribution, rather
than in the positions of the individual particles.
The probability of finding the system in a configuration with the
density $\rho(\vct{r},t)$ is given by
\begin{equation}
\label{eq:Pdensdef}
\Pdensity[\rho] =  \int 
\pdensity(\{\vct{r}_{\ell}\},t)\,
\delta\left[ \rho(\vct{r},t)-\mdensity(\vct{r},t;\{\vct{r}_i\})\right]
\rmd^3 r_1 \dots \rmd^3 r_N .
\end{equation}
$\mdensity(\vct{r},t;\{\vct{r}_i\})$ is a sum of $N$ $\delta$-functions and
therefore, from the above definition, $\Pdensity[\rho]$ can only be
non-zero when
$\rho(\vct{r},t)$ is also a sum of exactly $N$ $\delta$-functions.
In order to make a clear distinction between smooth densities and spiky
sums of delta functions we will keep a hat 
on the argument of $\Pdensity$\/.
The time evolution for $\Pdensity[\mdensity]$, which in fact is the
Fokker-Planck equation corresponding to \eref{eq:microevolution2}, is given by
\cite{frusawa00} (see also \cite{zinn-justin02} or appendix 8 in
\cite{goldenfeld92}):
\begin{equation}
\label{eq:microfp}
\fl
\frac{\partial \Pdensity[\mdensity]}{\partial t} = - \int
\frac{\delta }{\delta \mdensity(\vct{r},t)}\left\{
\grad \cdot \mdensity(\vct{r},t) \,\grad
\left[T\,\frac{\delta }{\delta \mdensity(\vct{r},t)} + \frac{\delta
\Hamil[\mdensity]}{\delta \mdensity(\vct{r},t)}
\right]\,\Pdensity[\mdensity]
\right\}\rmd^3 r
\end{equation}
with the functional $\Hamil[\mdensity]$ as given in
\eref{eq:hamil}\/. The derivation of \eref{eq:microfp} from the Langevin
equation \eref{eq:microevolution2} is path~5 connecting the two dotted
boxes in \fref{figure}\/.
Equivalently, one could also obtain \eref{eq:microfp} via a change of
variables from $\{\vct{r}_i\}_{i=1\dots N}$ in
\eref{eq:positionfp} to $\mdensity(\vct{r},t)$;
this is path 9 in \fref{figure}\/.
The equilibrium distribution corresponding to \eref{eq:microfp} is given by
$\Pdensity^{(eq)}[\mdensity] \propto
\exp\left[-\frac{1}{T}\,\left(\Hamil[\mdensity]
+\mu\,\mdensity\right)\right]$
with an arbitrary chemical potential $\mu$\/.
The freedom of choice for $\mu$ comes from the fact that
\eref{eq:microevolution} has the form of a conservation law.

Eq.\ \eref{eq:microevolution2} together with equation \eref{eq:hamil} is a DDFT
for $\mdensity(\vct{r},t)$, i.e.\ an equation of motion for the instantaneous
microscopic density. However, although equation \eref{eq:microevolution2} is
written in terms of $\mdensity(\vct{r},t)$, this sum of
delta functions contains the exact positions of all particles (modulo
permutations) and the solution of \eref{eq:microevolution2} is formally
equivalent to integrating \eref{eq:brownian} -- see also the appendix in
reference
\cite{kawasaki98}\/. Generally, one is not
interested in the individual trajectories of the particles, rather
in the time evolution of quantities (in particular the fluid density)
averaged over many realizations of the noise or, alternatively,
in the evolution of coarse grained quantities. 

\section{Ensemble averaged density}
\label{sec:ensemble}

We define the ensemble averaged density as the density $\mdensity(\vct{r},t)$
averaged over
many solutions of \eref{eq:microevolution} with equal initial conditions
but different realizations of the noise term $\bm{\xi}(\vct{r},t)$, i.e.
\begin{equation}
\adensity(\vct{r},t)= \int 
\mdensity(\vct{r},t;\{\vct{r}_i\})\,\pdensity(\{\vct{r}_i\},t)\,
\rmd^3 r_1\dots \rmd^3 r_N 
= \langle \mdensity(\vct{r},t) \rangle.
\end{equation}
In other words, in order to measure $\adensity(\vct{r},t)$ one solves 
\eref{eq:microevolution} many times for different realizations of the noise
$\bm{\xi}(\vct{r},t)$ and averages the result at each time $t$.
Since the particles are indistinguishable,
$\pdensity(\{\vct{r}_i\},t)$ is symmetric with respect to permutations of
particles and therefore $\adensity(\vct{r},t)= N \int
\pdensity(\vct{r},\vct{r}_2 ,\dots,\vct{r}_N,t)\,\rmd^3 r_2 \dots \rmd^3
r_N$\/. We can obtain the time evolution of $\adensity(\vct{r},t)$ by
integrating out $N-1$ degrees of freedom in \eref{eq:positionfp} to obtain
\cite{archer04,dhont96}:
\begin{equation}
\fl
\frac{\partial \adensity(\vct{r},t)}{\partial t}  = 
-\grad \cdot \left\{ \left[ \exforce(\vct{r},t) - T\,\grad\right]
\adensity(\vct{r},t) +\!\int\!\corrfunc(\vct{r},\vct{r}',t) \,
\intforce(\vct{r}-\vct{r}')\,\rmd^3 r'\right\},
\label{eq:firstbbgky}
\end{equation}
where
\begin{equation}
\corrfunc(\vct{r},\vct{r}',t) = N\,(N-1)\,\int
\pdensity(\vct{r},\vct{r}',\vct{r}_3,\dots , \vct{r}_N,t) \,\rmd^3 r_3\dots
\rmd^3 r_N,
\end{equation}
is the two-body density distribution function.
We have assumed that surface terms which appear in the
partial integrations vanish. Here we see that
for interacting particles, the time evolution of the one-body density
distribution $\adensity(\vct{r},t)$
depends upon the two-body density distribution function, i.e.\ on the
particle correlations in the system. In fact,
\eref{eq:firstbbgky} is the first equation of a hierarchy of equations for
the $n$-body density distribution functions (similar to the
BBGKY hierarchy) which one obtains
by integrating out $(N-n)$ degrees of freedom in \eref{eq:positionfp} -- see
\cite{archer04}. 

So far, everything is exact, but in order
to make use of this hierarchy one needs a closure. One way,
which is employed in reference \cite{archer04},
makes use of a result from equilibrium density
functional theory, in which
there is an exact sum rule connecting the equilibrium two-body density
distribution function $\eqcorrfunc(\vct{r},\vct{r}')$ to the gradient of the
one-body direct correlation function $\dcorr(\vct{r})$
\begin{equation}
k_BT \adensity(\vct{r})\,\grad \dcorr(\vct{r}) =
\int\!\eqcorrfunc(\vct{r},\vct{r}')\,\intforce(\vct{r}-\vct{r}')\,\rmd^3
r'.
\end{equation}
In equilibrium $\dcorr(\vct{r})$ is the effective one-body potential due to
interactions in the fluid and is given by the
functional derivative of $\freeex[\adensity]$, the excess (over ideal) part
of the Helmholtz free energy functional,
$-k_BT \dcorr(\vct{r})=\frac{\delta \freeex[\adensity]}{\delta \adensity}$
\cite{evans79,archer04}, and we obtain the following exact result
\begin{equation}
\adensity(\vct{r})\,\grad
\frac{\delta \freeex[\adensity]}{\delta \adensity(\vct{r})} = -
\int\!\eqcorrfunc(\vct{r},\vct{r}')\,\intforce(\vct{r}-\vct{r}')\,\rmd^3 r'.
\label{eq:sumrule}
\end{equation}
Assuming this relation remains valid
for the non-equilibrium fluid is equivalent to assuming that the
two-body correlations in the non-equilibrium fluid are the same as those in an
equilibrium fluid with the {\em same} one body density
profile \cite{marconi99,archer04}\/. 
Within this {\em approximation} we can write the time evolution
equation \eref{eq:firstbbgky} as \cite{marconi99,archer04}:
\begin{equation}
\frac{\partial \adensity(\vct{r},t)}{\partial t} = 
\grad\cdot \left[ \adensity(\vct{r},t)\,\grad \frac{\delta
\free[\adensity(\vct{r},t)]}{\delta \adensity(\vct{r},t)} \right],
\label{eq:averevol}
\end{equation}
with the Helmholtz free energy functional\footnote{Note that the equilibrium
excess Helmholtz free energy functional
$\freeex[\adensity]$, is only known exactly for the case of hard-rods in one
dimension \cite{percus76} (see also \cite{marconi99})\/. In practice one can
only obtain an approximation to
the exact functional. Some of the more recently derived functionals are
rather sophisticated and very accurate \cite{evans92}\/.}
\begin{equation}
\label{eq:freeenergy}
\fl
\free[\adensity(\vct{r},t)]
= \int \rho(\vct{r},t)\,\left[T\,\left(\ln \Lambda^3 \rho(\vct{r},t)
-1\right)
+\expot(\vct{r},t)\right] \,\rmd^3 r + \freeex[\adensity(\vct{r},t)].
\end{equation}
Since $\adensity(\vct{r},t)$ is a quantity averaged over the realizations
of the noise, the time evolution equation \eref{eq:averevol} is necessarily
deterministic. Eq.\ \eref{eq:averevol} was originally written down (without
derivation) by Evans \cite{evans79}\/.
The derivation of \eref{eq:averevol} presented in this
section corresponds to path~3 in \fref{figure}\/. In reference \cite{marconi99}
Marconi and Tarazona follow
path~4 in \fref{figure} and \eref{eq:averevol} is derived from
\eref{eq:microevolution} by performing an ensemble average.
Note that \eref{eq:averevol} also remains valid when the particles interact via
multi-body potentials \cite{archer04}.

\section{Coarse grained density}
\label{sec:coarse}

The ensemble averaged density $\adensity(\vct{r},t)$ discussed in 
\sref{sec:ensemble} is the quantity often considered by theorists,
but it is not the quantity one would measure by performing one
individual experiment (numerical or real)\/. In order to measure the density as
a function of time in an experiment one must define a probe volume in space
and a time window and count the number of particles in that probe volume
averaged over the time window.  Since we intend to resolve the atomistic
structure of the fluid, the spatial resolution has to be very good. 
In other words, we assume an experimental resolution function of the form 
$\kernel_1(\vct{r},t)=\kernel(t)\, \delta(\vct{r})$ and then the
measured density is
\begin{equation}
\label{eq:coarsedef}
\cdensity(\vct{r},t) = \int \kernel(t-t')\,\mdensity(\vct{r},t')\,\rmd t'.
\end{equation}
For simplicity we assume further that $\kernel(t)$ has a compact support.
Similarly, we can define a two-particle probe,
with a resolution function of the form
$\kernel_2(\vct{r},\vct{r}',t)=\kernel(t)\, \delta(\vct{r})\, \delta(\vct{r}')$.
The two-particle distribution function that we measure is
(c.f.\ equation~(A6) in \cite{evans79})
\begin{eqnarray} \fl
\label{eq:coarse2def}
\int \kernel(t-t')\,\mdensity(\vct{r},t')\,\mdensity(\vct{r}',t')\,\rmd t'
&=& \int \kernel(t-t')\,\sum_{i \neq j} \delta(\vct{r}_i(t')-\vct{r})
\delta(\vct{r}_j(t')-\vct{r}')\,\rmd t' \nonumber \\
&&+\int \kernel(t-t')\,\sum_i \delta(\vct{r}_i(t')-\vct{r})
\delta(\vct{r}_i(t')-\vct{r}')\,\rmd t' \nonumber \\
&=& \cdensity^{(2)}(\vct{r},\vct{r}',t)+\cdensity(\vct{r},t)\,
\delta(\vct{r}'-\vct{r}).
\end{eqnarray}
This equation defines the coarse grained two-particle density distribution
function, $\cdensity^{(2)}(\vct{r},\vct{r}',t)$\/. When the fluid is in
equilibrium, assuming ergodicity, and if the time window of
$\kernel(t)$ is large as compared to the time the
system needs to explore phase space locally, then
$\cdensity^{(2)}(\vct{r},\vct{r}') \approx \corrfunc(\vct{r},\vct{r}')$; we
therefore assume  $\cdensity^{(2)}(\vct{r},\vct{r}',t) \approx
\corrfunc(\vct{r},\vct{r}',t)$.
Here, locally means on the scale of the correlation length
and therefore this may not be true for long ranged correlations.

Using \eref{eq:microevolution} together with \eref{eq:coarsedef} and
\eref{eq:coarse2def}, 
the time evolution of $\cdensity(\vct{r},t)$ is given by
\begin{eqnarray}
\label{eq:coarse_1} \fl
\frac{\partial \cdensity(\vct{r},t)}{\partial t} = 
T\,\grad^2\cdensity(\vct{r},t) 
- \grad\cdot\cdensity(\vct{r},t)\,\exforce(\vct{r},t)
- \grad\cdot \!\int\!\cdensity^{(2)}(\vct{r},\vct{r}',t) \,
\intforce(\vct{r}-\vct{r}')\,\rmd^3 r'
\nonumber\\
+\grad\cdot \int\! \kernel(t-t')\,
\sqrt{\mdensity(\vct{r},t')}\,\bm{\xi}(\vct{r},t')\,\rmd  t'.
\end{eqnarray}
The delta function term in \eref{eq:coarse2def} leads to a term containing
$\intforce(0)=-\grad \intpot(r)|_{r=0}$,
namely the force between particles when the centers are
at zero separation. We assume
that this is zero. This will certainly be the case for particles interacting
via pair potentials of the form $\intpot(r)=\intpot_{hs}(r)+\intpot_{tail}(r)$,
where $\intpot_{hs}(r)$ is the hard-sphere pair potential.
Equation \eref{eq:coarse_1} is exact when $\exforce$ does not change over time
or when it only changes on time scales much longer than the support of
$\kernel(t)$. However, in this form equation
\eref{eq:coarse_1} is of little use, since the right hand
side involves $\mdensity(\vct{r},t)$ explicitly.
In the following we give a number of `hand-waving' arguments which allow
us to approximate the right hand side in terms of $\cdensity(\vct{r},t)$
only.  These arguments are far from being rigorous and 
are by no means a proper derivation of a DDFT for $\cdensity(\vct{r},t)$\/.
The reason for giving these arguments is that they lead to a stochastic
time evolution equation for $\cdensity(\vct{r},t)$ which has the same form
as the time evolution equation of $\adensity(\vct{r},t)$, i.e.\
\eref{eq:averevol}, but with an additional conserved noise term. In
reference
\cite{kawasaki94} Kawasaki uses a more formal spatial coarse graining
procedure and in reference \cite{perez-madrid2002} P\'erez-Madrid {\it
et.\ al.}~use a mesoscopic approach,
to obtain a similar result.

The time integration in the last term of equation \eref{eq:coarse_1} 
in principle leads to a non-Markovian time
evolution for $\cdensity(\vct{r},t)$\/. 
However, on time scales large as compared to the support of $K(t)$, the
dynamics of $\cdensity(\vct{r},t)$ should be approximately Markovian and
we write the last term  of equation \eref{eq:coarse_1} in the same form
as in \eref{eq:microevolution}, but with $\mdensity(\vct{r},t)$ replaced by
$\cdensity(\vct{r},t)$\/, so that we obtain a time evolution equation
identical to \eref{eq:firstbbgky} but with an additional multiplicative
noise term:
\begin{eqnarray}
\label{eq:firstbbgkynoise} \fl
\frac{\partial \cdensity(\vct{r},t)}{\partial t}  = 
-\grad \cdot \Bigg\{ \left[ \exforce(\vct{r},t) - T\,\grad\right]
\cdensity(\vct{r},t) +\!\int\!\cdensity^{(2)}(\vct{r},\vct{r}',t) \,
\intforce(\vct{r}-\vct{r}')\,\rmd^3 r'\nonumber\\
-\grad\cdot\sqrt{\cdensity(\vct{r},t)}\,\bm{\xi}(\vct{r},t)\Bigg\}.
\end{eqnarray}
Using the local equilibrium assumption applied in \sref{sec:ensemble},
we assume equation \eref{eq:sumrule} with $\adensity$ replaced by $\cdensity$
and $\adensity^{(2)}$ replaced by $\cdensity^{(2)}$ applies for the
non-equilibrium coarse grained density distributions, and
we substitute the term involving $\cdensity^{(2)}(\vct{r},\vct{r}',t)$
in \eref{eq:firstbbgkynoise} by a
term involving the functional derivative
of the excess part of the Helmholtz free energy functional, to obtain a
stochastic version of \eref{eq:averevol}:
\begin{equation}
\frac{\partial \cdensity(\vct{r},t)}{\partial t} = 
\grad\cdot \left[ \cdensity(\vct{r},t)\,\grad \frac{\delta
\free[\cdensity(\vct{r},t)]}{\delta \cdensity(\vct{r},t)} 
+ \sqrt{\cdensity(\vct{r},t)}\,\bm{\xi}(\vct{r},t)\right].
\label{eq:coarseevol}
\end{equation}
Note the similarity with the time evolution equation \eref{eq:microevolution2}
for the instantaneous microscopic density $\mdensity(\vct{r},t)$.
However, a significant difference is that it is
the Helmholtz free energy functional, $\free[\cdensity]$, given by equation
\eref{eq:freeenergy} with $\adensity$ replaced by $\cdensity$, that enters
into equation
\eref{eq:coarseevol}\footnote{More strictly, we expect the free energy
functional in \eref{eq:coarseevol} to be a coarse-grained free energy
functional, $\bar{\free}[\cdensity]$. We are making the approximation
$\bar{\free}[\cdensity] \simeq \free[\cdensity]$. See \cite{reguera04} for a
recent discussion of the difference between these free energy functionals.},
whereas $\Hamil[\mdensity]$ entering equation
\eref{eq:microevolution2} is simply the functional as defined
in \eref{eq:hamil}. Unfortunately, in reference \cite{frusawa00},
this distinction is not made clear.
In developing \eref{eq:coarseevol} we followed path~6 in \fref{figure}\/.
We can obtain the Fokker-Planck equation corresponding to equation
\eref{eq:coarseevol} by discretizing it in space.
Taking the continuum
limit of the Fokker-Planck equation for the discretized Langevin equation
\cite{munakata89,rauscher04a},
we obtain an equation very similar to \eref{eq:microfp}:
\begin{equation}
\label{eq:coarsefp}
\fl
\frac{\partial \Cdensity[\cdensity]}{\partial t} = - \int
\frac{\delta }{\delta \cdensity(\vct{r},t)}\left\{
\grad \cdot \cdensity(\vct{r},t) \,\grad
\left[T\,\frac{\delta }{\delta \cdensity(\vct{r},t)} + \frac{\delta
\free[\cdensity]}{\delta \cdensity(\vct{r},t)}
\right]\,\Cdensity[\cdensity]
\right\}\rmd^3 r.
\end{equation}
This calculation corresponds to path~8 in \fref{figure}\/.
The equilibrium distribution corresponding to \eref{eq:coarsefp} is given by
$\Cdensity^{(eq)}[\cdensity] \propto
\exp\left[-\frac{1}{T}\,\left(\free[\cdensity]
+\mu\,\cdensity\right)\right]$ with an arbitrary chemical potential $\mu$\/.
This result justifies {\em a posteriori} our, initially unjustified,
approximation for the noise term in \eref{eq:coarse_1}, which we
used to obtain \eref{eq:coarseevol}\/.

In this section we considered the temporally coarse grained density
$\cdensity(\vct{r},t)$ because we wished to make a comparison with
the DDFT obtained for the ensemble averaged density $\adensity(\vct{r},t)$
in the previous section, which is used to describe the time evolution of
the structure of the system at microscopic length scales. The derivation was by
no means rigorous. However, a
spatial coarse graining of \eref{eq:microfp} can also lead to an equation
of the same form as \eref{eq:coarsefp} but with a particular (approximate) form
for the Helmholtz free energy functional -- see reference
\cite{kawasaki94}\/. This last route follows path~7 in \fref{figure}\/.

\section{Discussion}
\label{sec:conclusion}

In \sref{sec:brown} and \sref{sec:coarse}
we derived stochastic DDFTs (Langevin equations)
for the instantaneous microscopic density $\mdensity(\vct{r},t)$, equation
\eref{eq:microevolution2}, and the coarse grained density
$\cdensity(\vct{r},t)$, equation \eref{eq:coarseevol}, respectively,
as well as the corresponding
Fokker-Planck equations. In \sref{sec:ensemble} we also derived a
deterministic DDFT, equation \eref{eq:averevol}, for the
ensemble (or noise) averaged density $\adensity(\vct{r},t)$\/. The
equations for the instantaneous microscopic density $\mdensity(\vct{r},t)$
are of little use because $\mdensity(\vct{r},t)$ directly encodes the
positions of all particles (modulo permutations); it is simpler in this case
to follow the particle trajectories directly by integrating equation
\eref{eq:brownian}\/. But, depending on the
situation one is trying to describe, it can be reasonable to consider either
the coarse grained density $\cdensity(\vct{r},t)$ or the ensemble
averaged density $\adensity(\vct{r},t)$\/. In fact, their time evolution
equations, \eref{eq:coarseevol} and \eref{eq:averevol} respectively, are
identical except for the additional noise term in \eref{eq:coarseevol}\/.
Both (approximate) equations
are consistent with equilibrium thermodynamics. While the equilibrium value
of $\adensity(\vct{r},t)$ (i.e.\ in the limit $t \rightarrow \infty$),
which we denote $\adensity(\vct{r})$, is correctly given by the stationary
point of the Helmholtz free energy functional $\frac{\delta
\free[\adensity]}{\delta \adensity(\vct{r})} = \mu$
\cite{marconi99,archer04,evans79}, the time evolution of the coarse
grained density $\cdensity(\vct{r},t)$ also includes fluctuations about the
equilibrium profile $\adensity(\vct{r})$\/. In other words
$\adensity(\vct{r})$ is the most probable density distribution in
$\Cdensity^{(eq)}[\cdensity]$\/.
For non-interacting particles, as pointed out in reference \cite{marconi99},
the deterministic DDFT in equation \eref{eq:averevol} reduces to the
(exact) drift-diffusion equation for Brownian
particles in an external potential.
These observations suggest that the DDFT provides a good approximation to the
exact time evolution of the particle density, particularly when the system is
either close to equilibrium or when the particles interact weakly.

We believe that the origin of the debate as to whether the DDFT should be a
stochastic or deterministic equation is a matter of confusion between the
three types of densities defined in this paper. In particular, the
distinction between the instantaneous microscopic density
$\mdensity(\vct{r},t)$ and the coarse grained density
$\cdensity(\vct{r},t)$ is of great importance, as noted also in reference
\cite{kawasaki98}\/. The confusion is most apparent in two recent
contributions to the field, references \cite{dean96,frusawa00}\/.  In both
papers, the Langevin and Fokker-Planck equation for $\mdensity(\vct{r},t)$,
namely \eref{eq:microevolution} and \eref{eq:microfp}, respectively, are
derived correctly, but then the microscopic instantaneous density
$\mdensity(\vct{r},t)$, which is a sum of delta functions, is confused with
the coarse grained density $\cdensity(\vct{r},t)$\/. In addition to this
confusion there is, in references \cite{dean96,frusawa00}, a lack of a clear
distinction between the
functional $\Hamil[\mdensity]$ and the free energy functional
$\free[\cdensity]$\/. Bearing in mind
the results from equilibrium density functional theory
\cite{evans79,evans92} this distinction should be evident immediately. 

The authors of reference \cite{frusawa00} argue that a DDFT
has to be stochastic in nature.
This misconception has its roots in
reference \cite{dean96} where the difference between $\mdensity(\vct{r},t)$ and
$\cdensity(\vct{r},t)$ is not recognized at all. This difference
is further obscured by the fact that the time
evolution equation for the {\em spatially} coarse grained density derived
in \cite{kawasaki94} involves a free energy functional whose form is strikingly
similar to $\Hamil[\mdensity]$ in \eref{eq:hamil}\/. However, the crucial
difference is that \eref{eq:hamil} contains the bare pair interaction
potential while the functional derived in \cite{kawasaki94} contains the
direct pair correlation function, instead of $V(r)$, and therefore leads to the
proper thermodynamic equilibrium state.
We wish to emphasize that a DDFT for $\adensity(\vct{r},t)$, the density
distribution averaged over all realizations of the stochastic noise, must be
deterministic.

In summary, one can justify both a stochastic as well as a deterministic
DDFT, depending on the quantities of interest. In this sense, there is a
confusion rather than a controversy about the nature of DDFT. 

\ack
We thank R.~Evans for fruitful discussions on the present work and 
AJA acknowledges the support of EPSRC under grant number GR/S28631/01.




\end{document}